\documentclass[11pt]{article}
\usepackage{moriond,epsfig}

\def\be{\begin{equation}}
\def\ee{\end{equation}}
\def\bea{\begin{eqnarray}}
\def\eea{\end{eqnarray}}

\def\lsim{\raise0.3ex\hbox{$\;<$\kern-0.75em\raise-1.1ex\hbox{$\sim\;$}}}
\def\gsim{\raise0.3ex\hbox{$\;>$\kern-0.75em\raise-1.1ex\hbox{$\sim\;$}}}
\begin{document}
\vspace*{4cm}
\title{SUSY: NEW PERSPECTIVES AND VARIANTS}

\author{C. MU\~NOZ}

\address{Departamento de F\'{\i}sica
      Te\'orica C-XI and Instituto de F\'{\i}sica
      Te\'orica C-XVI,\\ 
      Universidad Aut\'onoma de Madrid,
      Cantoblanco, 28049 Madrid, Spain.}

\maketitle\abstracts{
Although supersymmetry (SUSY) is thirty five years old, it is
still one of the most attractive theories for
physics beyond the standard model.
Assuming that SUSY will be discovered at the LHC, the key question is:
What SUSY model do we expect to be the correct one ?
After reviewing briefly the advantages and problems of SUSY, 
several interesting models
that have been proposed in the literature will be discussed.
In particular, models such as the MSSM, BRpV, NMSSM, and possible
extensions.
We will also introduce the $\mu\nu$SSM whose interest resides in the fact that
it generates a solution to the (famous) $\mu$ problem of SUSY models
that is connected to the (nowadays very popular) neutrino physics.
}

\section{Introduction}

We know from the past that symmetries are crucial in physics. The laws of modern
physics are invariant under certain symmetries: Invariance under
Lorentz
transformations is the origin of special relativity, and
invariance under local gauge transformations is the key point of
the Standard Model (SM), $SU(3)_c\times SU(2)_L\times U(1)_Y$.
Supersymmetry (SUSY) was proposed \cite{rusos,history} in the early 1970's
as an invariance of the theory under the interchange of fermions and
bosons. However,
knowns fermions like quarks, electron and neutrino, and bosons
like gluons, $W^\pm$ and photon are not married up in this fashion.
Instead, every known particle should have a (super) partner \cite{fayet}.
Thus the spectrum of elementary particles is doubled.
There are squarks and sleptons as superpartners of
quarks and leptons, 
and gluinos, Winos, Zino and photino as superpartners
of gluons,  $W^\pm , Z$ and photon.
Obviously, we have not detected SUSY particles with the same masses as
their SM partners (e.g. there is no selectron with mass $\sim 0.5$ MeV),
and actually 
accelerator physics imposes important lower bounds on their masses.
This also implies that SUSY cannot be an exact symmetry of Nature and should
be broken.

The SM particles together with their massive SUSY partners
constitute the so-called
Minimal Supersymmetric Standard Model \cite{mssm} (MSSM).
Since the Higgs has a fermionic superpartner, the Higgsino, two
Higgs doublets with opposite hypercharges, $H_1$ and $H_2$, 
must be present in order 
to avoid anomalies. 
In addition, because of the structure of SUSY, these two Higgs doublet
superfields are necessary to give masses to all quarks and leptons.
Clearly the MSSM has potentially a 
very rich phenomenology. Let us also remark that
the joining of the three gauge coupling constants 
at a single unification scale
agrees with the LEP experimental results. 
Whether this is a hint or just a coincidence, will be clarified
by the LHC.

In any case, 
despite the absence of experimental verification, relevant theoretical
arguments can be given in favour of SUSY. 
First of all, SUSY solves the so-called gauge hierarchy problem.
If we believe that the SM should be embedded within a 
Grand Unified Theory (GUT) with a typical scale
$M_{GUT}\sim 10^{15}$ GeV or within a 
more fundamental theory
including gravity with a characteristic scale
$M_{Planck} \sim 10^{19} $ GeV,
then we are faced with the hierarchy problem. There is no symmetry protecting
the masses of the scalar particles 
against quadratic divergences in perturbation
theory. Therefore they will be proportional to the huge cut-off scale 
$\sim M_{GUT}$ or $M_P$. 
The Higgs particle is included in the SM because of its good
properties: it can have a vacuum expectation value (VEV) without breaking 
Lorentz invariance, inducing the spontaneous breaking of the
electroweak (EW)
symmetry at the same time that generates the gauge boson masses, and
the 
fermion masses through Yukawa
couplings. But, of course, all these properties are due to the fact that
the Higgs is a scalar particle. As mentioned above, this leads to a huge
mass for it and, 
as a consequence of the minimization, for the $W$ and $Z$ gauge bosons.
This problem of naturalness, 
to stabilize $M_W \ll M_{GUT}, M_P$ against quantum corrections,
is solved in SUSY since 
now the masses of the scalars and the masses of their partners, 
the fermions, are related. As a consequence, only a logarithmic divergence
in the Higgs scalar mass is left. 
In diagrammatic language, the dangerous diagrams
of SM particles are cancelled with new ones which are present due to the
existence of the additional partners and couplings. 

Nevertheless, let us recall that SUSY was not invented to solve
this hierarchy problem. As mentioned above, it was created as a
new kind of symmetry which relates bosons and fermions. 
Notice that the latter also involves that the Higgs is no longer a mysterious particle
as it stands in the SM: the only fundamental scalar particle
which exists. Now, the Supersymmetric
Standard Model (SSM) is naturally
plenty of fundamental scalars (squarks, sleptons and Higgses) related through
SUSY with their fermionic partners (quarks, leptons and Higgsinos). 

Let us also mention that SUSY allows us to understand better
how the EW symmetry, $SU(2)_L\times U(1)_Y$, is broken.
Whereas in the case of the SM the quadratic and quartic terms of
the Higgs potential, $V=m^2 H^*H  + \lambda (H^*H)^2 $, with $m^2 <0$, have
to be 
postulated "ad hoc",
in the context of the SSM
they appear in a 
natural way.
The quartic terms arise from the usual $D$-term contributions
and $\lambda$ is given by the EW coupling constants,
and the quadratic terms arise once SUSY is broken and
masses are generated for all scalar particles.
Fortunately, these mass terms are `soft' in the sense that they
do not induce quadratic divergences, and therefore do not
spoil the SUSY solution to the gauge hierarchy problem.
In addition, the large top mass produces radiatively a 
negative mass square for the Higgs $H_2$ inducing
 the breaking 
$SU(2)_L\times U(1)_Y\to U(1)_{em}$.

Besides, we should not forget that the 
local version of SUSY leads to a partial unification of the SM with
gravity, the so-called Supergravity
(SUGRA). 
In this theory the graviton has a SUSY partner, the
gravitino,
and actually the breaking of SUGRA in order to generate a mass for
the gravitino is an interesting mechanism
to generate also the (soft) masses of the superpartners of the 
SM \cite{corfu}.
Last but not least, SUSY seems to be a crucial ingredient of string theory.

Another (more model dependent) advantage of SUSY is related to the issue of 
the existence of dark matter in the Universe. 
If the superpotential of the theory
conserves a discrete symmetry called 
$R$-parity (+1 for particles and -1 for superpartners),
SUSY particles are produced 
or destroyed only in pairs and, as a consequence, the lightest 
supersymmetric particle (LSP) 
is absolutely stable.
This implies that the LSP 
is a possible candidate for dark matter. 
Concerning this point, it is remarkable that in most of the parameter space
of the MSSM
the LSP is the lightest neutralino, 
a physical superposition of the Bino, and neutral Wino and Higgsinos.
The neutralino is
obviously 
an electrically neutral 
(also with no strong interactions) particle, and 
this is welcome since otherwise the LSP 
would bind to nuclei and would be excluded as a candidate
for dark matter from unsuccessful searches for exotic heavy 
isotopes \cite{isotopes}.
Therefore, in the MSSM, typically the lightest neutralino 
is a very good
dark matter candidate \cite{reviewmio}.
Actually, the neutralino is a Weakly Interacting Massive Particle (WIMP), and
therefore it is able to produce in some regions of the 
SUSY parameter space a value of the relic
density
of the correct order of magnitude, $\Omega_{DM}h^2\sim 0.1$.
Let us finally remark that the fact that the LSP is stable,
and typically neutral, implies implies that
a major signature in accelerator experiments 
for $R$-parity conserving models is represented by events with missing 
energy.

On the other hand, 
there is no a fundamental reason to impose $R$-parity conservation
in SUSY models. 
Actually, lepton and baryon number violating terms in the superpotential
like 
$\epsilon_{ab} \left(\lambda^{ijk} \hat L_i^a \hat L_j^b \hat e^c_k 
+ 
\lambda'^{ijk} \hat L_i^a \hat Q_j^b \hat d^c_k 
+ 
\mu^i  \hat L_i^a\hat H_2^b \right)$, 
and
$\lambda''^{ijk} \hat d^c_i \hat d^c_j \hat u^c_k$,
respectively, with
$i,j=1,2,3$ generation indices and 
$a,b=1,2$ 
$SU(2)$ indices,
which break explicitly $R$-parity, are in principle allowed by gauge
invariance. As is well known, to avoid too fast proton decay mediated by
the exchange of squarks of masses of order the EW scale,
the presence together of terms of the type  
$\hat L \hat Q \hat d^c$ and 
$\hat d^c \hat d^c \hat u^c$ must be forbidden, unless
we impose very stringent bounds such as e.g.
$\lambda'^*_{112}\dot \lambda''_{112} \lsim 2\times 10^{-27}$.
However the latter values for the couplings are not very natural, and
for constructing viable SUSY models one usually 
forbids at least one of the operators
$LQd^c$ or $u^cd^cd^c$. 
The other type of operators above are not so stringently supressed, and
therefore still a lot of freedom remains \cite{dreiner}.

There is a large number of works in the 
literature \cite{Barbier} exploring the possibility
of $R$-parity breaking in SUSY models, and its consequences
for the detection of SUSY at the LHC \cite{dreiner2}. 
A popular model is the so-called Bilinear $R$-parity Violation (BRpV)
model \cite{hall},
where bilinear terms of the above type, $\hat L \hat H_2$, 
are added to the MSSM.
In this way it is in principle possible to generate neutrino masses 
withouth including in the model 
right-handed neutrinos, unlike the MSSM.
Analyses of mass matrices\cite{mass} in the BRpV, as well as 
studies of signals at accelerators \cite{signals} 
has been extensively carried out
in the literature.
Other interesting models 
are those producing the spontaneous breaking of $R$-parity
through the VEVs of singlet fields \cite{vallee}.
 
Of course, the 
phenomenology of models where $R$-parity is broken
is going to be very different from that of models where $R$-parity is 
conserved. Needless to mention, the LSP is no longer stable, and
therefore not all SUSY chains must yield
missing energy events at colliders.
Obviosly, in this context
the neutralino is no longer a candidate for dark matter.
Nevertheless, other candidates can be found in the literature,
such as the
gravitino \cite{yamaguchi}, 
the well-known axion, 
and many other (exotic) particles \cite{reviewmio}. 

Up to now
we have mentioned several advantages of SUSY, but it is fair to say
that
problems are also present in this theory.
For example, it is not clear yet the mechanism of SUSY breaking
generating the soft masses for scalar particles of order the EW scale. As
mentioned above, dangerous baryon and lepton number violation operators
may be present, and they must be supressed by some mechanism.
Dangerous charge and colour breaking minima may also be present
in the parameter space of SUSY models \cite{color}.
There is the possibility of having too large contributions to 
Flavour Changing Neutral Currents
(FCNC), as well as to the Neutron Electric Dipole Moment
(EDM). There might be a fine-tuning problem in SUSY models.
Finally, there is also the so-called $\mu$ problem \cite{mupb}, arising
from the presence of a mass term for the Higgs fields in the
superpotential $\mu \hat H_1 \hat H_2$.

Concerning the latter, the
Next-to-Minimal Supersymmetric Standard Model \cite{nmssm} (NMSSM),
provides an elegant solution via the introduction of a singlet
superfield $\hat S$. 
In the simplest form of the superpotential, which is scale invariant
and contains the $\hat S \hat H_1 \hat H_2$ coupling, 
an effective $\mu$ term is generated when the scalar component
of $\hat S$ acquires a VEV of order the SUSY breaking scale. 
The NMSSM 
has been extensively analysed in the literature \cite{nmssm,nmssm5,nmssm2},
as well as
its possible extensions generating neutrino masses.
In Sects.~2 and 3 
we will review the MSSM, and the NMSSM, respectively,
and their extensions.


Let us finally mention that a model,
the so-called $\mu\nu SSM$, has been proposed \cite{mulo2,mulo}
to solve the
$\mu$ problem of the MSSM without having to introduce
an extra singlet superfield as in the NMSSM. 
Interestingly, the solution is connected to the neutrino physics.
In particular, terms of the type $\nu^c H_1H_2$ in the superpotential 
generate the $\mu$ term spontaneously through right-handed sneutrino 
VEVs.
In addition, terms of the type  $(\nu^c)^3$ forbid
a global $U(1)$ symmetry in the superpotential,  
avoiding therefore the existence of a Goldstone boson.
Besides, the latter contribute to generate
effective Majorana masses for neutrinos at the EW scale.
On the other hand, these two type of terms break lepton number and $R$-parity 
explicitly implying that
the phenomenology of this model is very different from the one 
of the MSSM/NMSSM.
For example, the usual neutralinos are now mixed with the neutrinos.
Since we have a generalized see-saw mechanism at the EW scale,
for a Dirac mass of the heaviest neutrino of order the mass
of the electron, $0.1$ MeV, an eigenvalue reproducing 
the correct scale of the heaviest neutrino
mass, 0.01 eV, is obtained. Playing with the hierarchies in the Dirac masses
one can obtain the other neutrino masses.
In Sect.~4 we will review the $\mu\nu SSM$. 

We have mentioned in this Introduction several interesting SUSY
models, but let us remark 
that others have been proposed during the years, and therefore our
list is by no means complete.
In any case, 
pretty soon, at the end of this year 2007, 
the LHC will start operations, thus the
crucial question is by now: What SUSY model do we expect to be
discovered?
The reader can find in the rest of the paper several of the above
models discussed in some detail, and pick up her/his preferred one.

\section{MSSM and extensions}

The MSSM is the most popular SUSY extension of the SM.
It was the first SUSY model studied in detail in the
literature, 
and
it is the simplest extension: no extra fields are
included apart from the SUSY partners of the SM fields. 
Let us discuss briefly the structure of the MSSM.

Working for the moment with massless neutrinos,
in addition to the Yukawa couplings for quarks and charged leptons, the
MSSM superpotential contains the so-called $\mu$-term
involving the Higgs doublet
superfields, $\hat H_1$ and $\hat H_2$,
\begin{equation}
\begin{array}{rcl}
W & = &
\ \epsilon_{ab} \left(
Y_u^{ij} \, \hat H_2^b\, \hat Q^a_i \, \hat u_j^c +
Y_d^{ij} \, \hat H_1^a\, \hat Q^b_i \, \hat d_j^c +
Y_e^{ij} \, \hat H_1^a\, \hat L^b_i \, \hat e_j^c \right)
- \epsilon_{ab}\ \mu  \hat H_1^a \hat H_2^b\ ,
\end{array}\label{superpotentialmssm}
\end{equation}
where we take $\hat H_1^T=(\hat H_1^0, \hat H_1^-)$, $\hat H_2^T=(\hat
H_2^+, \hat H_2^0)$, $\hat Q_i^T=(\hat u_i, \hat d_i)$, 
$\hat L_i^T=(\hat \nu_i, \hat e_i)$, $i,j=1,2,3$ are generation indices, 
$a,b$ are
$SU(2)$ indices, and $\epsilon_{12}=1$.
It is worth noticing that this superpotential conserves
$R$-parity.
The presence of the $\mu$-term  
in (\ref{superpotentialmssm}) is essential to avoid 
the appearance of an unacceptable Goldstone boson  
associated to a global $U(1)$
symmetry $H_{1,2}\to e^{i\alpha}H_{1,2}$,
when this is broken by the VEVs of the Higgses giving
masses to all quarks and leptons.
In addition, the minimum of the Higgs potential without including this
term
occurs for a vanishing VEV for $H_1$, and therefore $d$-type quarks
and $e$-type leptons remain massless.
Unfortunately, the $\mu$-term  introduces a naturalness problem,
the so-called $\mu$ problem \cite{mupb}. Note that, to this respect,
the $\mu$-term is purely supersymmetric, and therefore the natural
scale of $\mu$ would be $M_{GUT}$ or $M_{Planck}$.
Thus, any complete explanation of the EW scale must
justify the origin of $\mu$, i.e. why its value is of order $M_W$ and
not $M_{GUT}$
or $M_P$.

This problem has been considered by several authors and different
possible solutions have been proposed, producing an 
effective $\mu$-term \cite{musolutions,musolutions2,musolutions3}.
On the other hand, there are also very interesting 
solutions in the literature that
necessarily introduce new structure beyond the MSSM at low energies. 
Several of these
solutions, and the associated SUSY models, will be discussed in the
next sections.

Concerning the spectrum of the MSSM,
as is well known, after the EW breaking the model is left with
three neutral Higgses and two charged ones. 
Although at tree level the mass of the lightest Higgs, $m_h$, is 
bounded by $M_Z$, loop corrections allow a larger bound, $m_h\lsim 135$
GeV,
still fulfilling the experimental constraint $m_h\gsim 114$ GeV.
In the MSSM 
there are also four neutralinos, 
the physical 
superpositions of the Bino, and neutral Wino and Higgsinos.
As discussed in the Introduction, the lightest one is a good candidate for
dark matter.


\subsection{Neutrino masses}


Neutrino experiments have confirmed during the last years 
that neutrinos are massive \cite{experiments}.
As a consequence, all theoretical models must be modified in order to
reproduce this result.
In particular, 
it is natural in the context of the MSSM 
to supplement the ordinary  neutrino
superfields, $\hat \nu_i$, 
contained in the $SU(2)_L$-doublet, $\hat L_i$, with
gauge-singlet neutrino superfields, $\hat \nu^c_i$.
Thus, 
in addition to the usual Yukawa couplings for quarks and charged
leptons, 
and the $\mu$-term, the
superpotential (\ref{superpotentialmssm}) 
may contain new terms such as Yukawa couplings for neutrinos, and possible
Majorana mass terms:
\begin{equation}
\begin{array}{rcl}
\delta W & = &
\ \epsilon_{ab} 
Y_\nu^{ij} \, \hat H_2^b\, \hat L^a_i \, \hat \nu^c_j 
+ m_M^{ij} \hat \nu^c_i\hat \nu^c_j\ .
\end{array}\label{deltasuperpotentialMajorana}
\end{equation}
%
Clearly, one has that the 
couplings $Y_{\nu}$ 
determine the Dirac masses for the 
neutrinos,
$m_D = Y_{\nu}v_2$, whereas $m_M$ are the Majorana masses.
If $m_D<<m_M$, both type of contributions induce very 
light neutrino masses of order
%
\begin{equation}
m_\nu\sim 
\frac{m_D^2}{m_M}\ .
\label{neutrinomass}
\end{equation}
This is the SUSY extension \cite{seesawsusy} 
of the well-known see-saw mechanism 
studied in the context of the SM \cite{seesawsm}.
Notice also that there will be 
a heavy neutrino with mass of order $m_M$.
Result (\ref{neutrinomass}) was considered very interesting in
the context of the early attempts to connect the standard model with
GUTs where $M_{GUT}\sim 10^{15}$ GeV.
Indeed, since $v_2\sim 10^2$ GeV, 
with a neutrino Yukawa coupling of order one, 
and $m_M\sim10^{15}$ GeV,
one is able
to obtain neutrino masses as small as $10^{-2}$ eV.
Of course, 
this can be consider an improvement with respect to the use of a 
purely Dirac mass for the neutrino, which  
would imply the necessity of explaining 
a Yukawa coupling of order $10^{-13}$,
i.e. thirteen orders of magnitude smaller than the one that we need
with a GUT-scale see-saw.
Recall that one chooses to forbid couplings with such small values,
when discussing R-parity violating couplings producing proton decay,
because they seem not very natural.

Nevertheless, let us remark  
that a see-saw at the EW scale 
is also a very interesting possibility.
Since we know that the Yukawa coupling of the electron has to be
of order $10^{-6}$, why the one of the neutrino should be
six orders of magnitude larger?
With a EW-scale see-saw, i.e. $m_M\sim 1$ TeV, 
a neutrino Yukawa coupling of order of the one of the electron
generating $m_D\sim10^{-4}$ GeV, is
sufficient to produce a neutrino mass of order $10^{-2}$ eV
(see eq.~(\ref{neutrinomass})).
This possibility is also interesting because
one does not need to introduce in the game any ad-hoc
high-energy scale.
It is worth mentioning here that in some string constructions, where
SUSY standard-like models can be obtained without the necessity
of a GUT, and
Yukawa couplings can be explicitly computed, those for neutrinos
cannot be as small as $10^{-13}$, and therefore the presence of a see-saw
at the EW scale is helpful \cite{AbelMunoz}.



On the other hand, it is worth remarking that
neutrino masses can also be obtained without using 
the singlet superfields $\hat \nu^c_i$. Adding to the superpotential 
(\ref{superpotentialmssm})
the bilinear terms
\begin{equation}
\begin{array}{rcl}
\delta W & = &
 \epsilon_{ab}\ \mu^i  \hat H_2^b \hat L_i^a
\  ,
\end{array}\label{bilinear2}
\end{equation}
neutrino masses are induced
through the mixing with
the neutralinos (actually only 
one mass at tree level and the other two at one loop).
The above terms break R-parity explicitly, as discussed in the
Introduction, and together
with  the superpotential 
(\ref{superpotentialmssm}) they
constitute the
BRpV.  
Although this is an interesting mechanism for generating neutrino masses,
notice that
the $\mu$ problem is augmented
with the three new bilinear terms.

\section{NMSSM and extensions}

The NMSSM 
provides an elegant solution to the $\mu$ problem of the MSSM via the
introduction of a singlet superfield $\hat S$ under the SM gauge group. 
The simplest form of the
superpotential, which is scale invariant, is given by:
\begin{equation}
\begin{array}{rcl}
W & = &
\ \epsilon_{ab} \left(
Y_u^{ij} \, \hat H_2^b\, \hat Q^a_i \, \hat u_j^c +
Y_d^{ij} \, \hat H_1^a\, \hat Q^b_i \, \hat d_j^c +
Y_e^{ij} \, \hat H_1^a\, \hat L^b_i \, \hat e_j^c \right)
- \epsilon_{ab}\ \lambda \hat S \hat H_1^a \hat H_2^b\ + \frac{1}{3}k \hat S\hat
S\hat S\ .
\end{array}\label{superpotentialnmssm}
\end{equation}
In this model, the usual
$\mu$ term is absent from the superpotential, and only dimensionless trilinear
couplings are present in $W$.
For this to happen it is usually invoked a $Z_3$ symmetry. 
On the other hand,
let us recall that this is actually what happens 
in the low-energy limit of string constructions,
where all fields are massless, and, as as consequence, 
only trilinear couplings are present in the superpotential.
Since string theory seems to be relevant for the unification of
all interactions, including gravity, this argument \cite{musolutions2} 
in favour of the
absence of a bare $\mu$ term in this kind of 
superpotentials 
is robust.

When the scalar component of the superfield $\hat S$,
denoted by $S$, acquires
a VEV of order the SUSY breaking scale, 
an effective interaction $\mu \hat H_1 \hat H_2$ is generated
through the fourth term in  (\ref{superpotentialnmssm}), with  
$\mu\equiv\lambda \langle S\rangle$.
This
effective coupling is naturally of order the EW scale if the SUSY  breaking
scale is not too large compared with $M_W$. In fact, the NMSSM is the simplest
SUSY extension of the SM in which the EW scale
exclusively originates from the SUSY breaking scale. 
The last term in (\ref{superpotentialnmssm}) 
is allowed by all symmetries,
and avoids, as the $\mu$-term in the MSSM, 
the presence of a Goldstone boson, in this case associated to the global $U(1)$
symmetry $H_1 H_2 \to e^{i\alpha}H_1 H_2$, $S\to e^{-i\alpha} S$.

In addition to the MSSM fields, the NMSSM contains an extra CP-even and CP-odd
neutral Higgs bosons, as well as one additional neutralino. These new fields
mix with the corresponding MSSM ones, giving rise to a richer and more complex
phenomenology.
For example, a 
light neutralino may be present. 
The mass of the lightest neutral Higgs state can be raised
arbitrarily by increasing the value of the new Higgs self-coupling
parameter $\lambda$.
Imposing that the coupling remains perturbative up to the 
Planck scale, then
the upper bound is larger than in the MSSM, $m_h\lsim 150$ GeV. 
Moreover, a very
light Higgs boson is not experimentally excluded. 
In particular, such a particle has a significant singlet composition,
thus scaping detection and being in agreement with accelerator data. 
The latter may modify the results concerning the 
possible detection of neutralino dark matter 
with respect to those of the MSSM \cite{nmssm2,belanger}.

Let us remark that
the superpotential (\ref{superpotentialnmssm}) 
has a $Z_3$ symmetry.
Therefore, one expects to have a 
cosmological domain wall problem \cite{wall,wall2} in this model. 
Nevertheless, 
there is a solution to this problem \cite{nowall}: 
non-renormalizable operators \cite{wall} in the superpotential can break 
explicitly the dangerous $Z_3$ symmetry, lifting the degeneracy of the 
three original vacua, and this can be done without introducing hierarchy 
problems. In addition, these operators can be chosen small enough as 
not to alter the low-energy phenomenology. 
An alternative solution 
\cite{langacker1,langacker2,langacker3,langacker4} 
uses an extra $U(1)$. 
Gauge invariance of 
$W$ under the new $U(1)$ forbids not only the $\mu$-term, but also 
the term $\hat S \hat S \hat S$, and thus the model
is free from the domain wall problem. Notice that
the Goldstone boson 
is eaten by the extra $Z$.
The extra $U(1)$ can also be very useful to forbid
R-parity violating terms producing proton decay \cite{langacker5,langacker6}.
Unless some quark and lepton Yukawa couplings are forbidden at 
tree level \cite{langacker4},
cancellation of anomalies with the new $U(1)$ requires the
introduction of new fermions charged under the SM group \footnote{We
thank J.R. Espinosa for useful comments about this point.}.
See ref.~\cite{barger} for a brief review of other variants of this
type of models.


\subsection{Neutrino masses}

Other extensions of the NMSSM, which in addition can help
us to understand
the origin of neutrino masses, can be considered. 
As in the case of the MSSM, one could add in principle the terms in eq.~(\ref{deltasuperpotentialMajorana}) generating a GUT scale see-saw.
But another 
interesting extension of the superpotential
(\ref{superpotentialnmssm}) generating dynamically an EW see-saw can be \cite{kitano}:
\begin{equation}
\begin{array}{rcl}
\delta W & = &
\ \epsilon_{ab} 
Y_\nu^{ij} \, \hat H_2^b\, \hat L^a_i \, \hat \nu^c_j 
+ k^{ij} \hat S \hat \nu^c_i\hat \nu^c_j\ .
\end{array}\label{deltasuperpotentialMajorana2}
\end{equation}
Here Majorana masses for neutrinos are generated
through the VEV of the singlet $S$. 
A similar superpotential was proposed in ref.~\cite{AbelMunoz}
including three families of Higgses, which may be naturally obtained
in the context of string constructions.

Inspired by the BRpV, another possibility consists of extending the superpotential (\ref{superpotentialnmssm}) with the bilinear 
terms in (\ref{bilinear2}) \cite{abada}.
Notice, however, that this strategy reintroduces 
the $\mu$ problem in a model which was supposed to solve it with the singlet, through the three new bilinear terms.

\section{$\mu\nu$SSM}

As discussed in the context of the MSSM and NMSSM,
experiments may induce us to introduce
gauge-singlet neutrino superfields, $\hat \nu^c_i$.
Now, given the
fact
that sneutrinos are allowed to get VEVs,
we may wonder why not to use terms
of the type $\hat \nu^c \hat H_1\hat H_2$ 
to produce an effective  $\mu$ term.
This would allow us to solve the $\mu$ problem of the MSSM, 
without having to introduce an extra singlet superfield
as in case of the NMSSM.
Thus the aim of what follows is to analyse the ``$\mu$ from $\nu$''
Supersymmetric Standard Model ($\mu$$\nu$SSM)  arising from
this proposal \cite{mulo2,mulo}: natural particle content without $\mu$ problem.



In addition to the MSSM Yukawa couplings for quarks and charged leptons, the
$\mu$$\nu$SSM superpotential contains Yukawa couplings for neutrinos, and  
two additional type of terms involving the Higgs doublet
superfields, $\hat H_1$ and $\hat H_2$, and the three
neutrino superfields, $\hat \nu^c_i$,
%
\begin{equation}
\begin{array}{rcl}
W & = &
\ \epsilon_{ab} \left(
Y_u^{ij} \, \hat H_2^b\, \hat Q^a_i \, \hat u_j^c +
Y_d^{ij} \, \hat H_1^a\, \hat Q^b_i \, \hat d_j^c +
Y_e^{ij} \, \hat H_1^a\, \hat L^b_i \, \hat e_j^c +
Y_\nu^{ij} \, \hat H_2^b\, \hat L^a_i \, \hat \nu^c_j 
\right)
\nonumber\\
& &   -  \epsilon_{ab} \lambda^{i} \, \hat \nu^c_i\,\hat H_1^a \hat H_2^b
+
\frac{1}{3}
\kappa^{ijk} 
\hat \nu^c_i\hat \nu^c_j\hat \nu^c_k\,.
\end{array}\label{superpotentialmunussm}
\end{equation}
%
In this model, the usual MSSM bilinear
$\mu$-term is absent from the superpotential, and only dimensionless trilinear
couplings are present in $W$. As argued in the previous section,
this is a natural situation in string constructions.
When the scalar components of the superfields $\hat\nu^c_i$,
denoted by $\tilde\nu^c_i$, acquire
VEVs of order the EW scale, 
an effective interaction $\mu \hat H_1 \hat H_2$ is generated
through the fifth term in  (\ref{superpotentialmunussm}), with  
$\mu\equiv
\lambda^i \langle \tilde \nu^c_i \rangle$.
The last type of terms in (\ref{superpotentialmunussm}) 
is allowed by all symmetries,
and avoids the presence of a Goldstone boson associated to a global $U(1)$
symmetry, similarly to the case of the NMSSM.
In addition, it contributes to generate 
effective Majorana masses for neutrinos at
the EW scale.
These two type of terms replace the two NMSSM terms
$\hat S\hat H_1 \hat H_2$ and
$\hat S\hat S \hat S$. 

It is worth noticing that these terms break explicitly
lepton number, and therefore, after spontaneous symmetry breaking,
a massless Goldstone boson (Majoron)
does not appear. On the other hand,
$R$-parity 
is also explicitly broken and this means that the 
phenomenology of the $\mu$$\nu$SSM is going to be very different from the one 
of the MSSM.
It is also interesting to realise that the Yukawa couplings producing 
Dirac
masses for neutrinos, the fourth term in (\ref{superpotentialmunussm}),
generate through the VEVs of $\tilde\nu^c_i$, 
three effective bilinear terms
$\hat H_2 \hat L_i$.
As mentioned above these
characterize the BRpV.
Let us finally mention that the terms 
$\nu^c H_1 H_2$ and $\nu^c\nu^c\nu^c$
have also been analysed as sources of the observed baryon asymmetry
in the Universe \cite{vallle}, and of neutrino masses and bilarge 
mixing \cite{sri}, respectively.

Notice that
the superpotential (\ref{superpotentialmunussm}) 
has a $Z_3$ symmetry, just like 
the NMSSM.
Therefore, one expects to have also a 
cosmological domain wall problem in this model.  
Nevertheless, 
the usual solutions to this problem discussed for the NMSSM above,
will also work in this 
case.


Working in the framework of SUGRA,
we will discuss now in more detail the phenomenology of the $\mu$$\nu$SSM.
Let us write first the soft terms appearing in the Lagrangian,
$\mathcal{L}_{
{soft}}$, after SUSY breaking, which in our conventions is
given by

\begin{equation}
\begin{array}{rcl}
-\mathcal{L}_{
{soft}} & =&
 (m_{\tilde{Q}}^2)^{ij} \, \tilde{Q^a_i}^* \, \tilde{Q^a_j}
+(m_{\tilde u^c}^{2})^{ij} 
\, \tilde{u^c_i}^* \, \tilde u^c_j
+(m_{\tilde d^c}^2)^{ij} \, \tilde{d^c_i}^* \, \tilde d^c_j
+(m_{\tilde{L}}^2)^{ij} \, \tilde{L^a_i}^* \, \tilde{L^a_j}
+(m_{\tilde e^c}^2)^{ij} \, \tilde{e^c_i}^* \, \tilde e^c_j
\nonumber \\
& & +
m_{H_1}^2 \,{H^a_1}^*\,H^a_1 + m_{H_2}^2 \,{H^a_2}^* H^a_2 +
(m_{\tilde\nu^c}^2)^{ij} \,\tilde{{\nu}^c_i}^* \tilde\nu^c_j 
\nonumber \\
& & +
\epsilon_{ab} \left[
(A_uY_u)^{ij} \, H_2^b\, \tilde Q^a_i \, \tilde u_j^c +
(A_dY_d)^{ij} \, H_1^a\, \tilde Q^b_i \, \tilde d_j^c +
(A_eY_e)^{ij} \, H_1^a\, \tilde L^b_i \, \tilde e_j^c 
\right.
\nonumber \\
& & +
\left.
(A_{\nu}Y_{\nu})^{ij} \, H_2^b\, \tilde L^a_i \, \tilde \nu^c_j 
+ 
{H.c.}
\right] 
\nonumber \\
& & +
\left[-\epsilon_{ab} (A_{\lambda}\lambda)^{i} \, \tilde \nu^c_i\, H_1^a  H_2^b
+
\frac{1}{3}
(A_{\kappa}\kappa)^{ijk} 
\tilde \nu^c_i \tilde \nu^c_j \tilde \nu^c_k\
+ 
{H.c.} \right]
\nonumber \\
& & -  \frac{1}{2}\, \left(M_3\, \tilde\lambda_3\, \tilde\lambda_3+M_2\,
  \tilde\lambda_2\, \tilde
\lambda_2
+M_1\, \tilde\lambda_1 \, \tilde\lambda_1 + 
{H.c.} \right) \,.
\end{array}
\label{2:Vsoft}
\end{equation}
%
In addition to terms from $\mathcal{L}_{
{soft}}$, the 
tree-level scalar potential receives the usual $D$ and $F$ term
contributions.
Once the EW symmetry is spontaneously broken, the neutral
scalars develop
in general
the following VEVs:
\begin{equation}\label{2:vevs}
\langle H_1^0 \rangle = v_1 \, , \quad
\langle H_2^0 \rangle = v_2 \, , \quad
\langle \tilde \nu_i \rangle = \nu_i \, , \quad
\langle \tilde \nu_i^c \rangle = \nu_i^c \,.
\end{equation}
%
In what follows it will be enough for our purposes to neglect mixing
between generations in (\ref{superpotentialmunussm}) and (\ref{2:Vsoft}), and
to assume that only one generation of
sneutrinos gets VEVs,
$\nu$, $\nu^c$. The extension of the analysis to all generations
is straightforward, and the conclusions are similar.
We
then obtain for the tree-level neutral scalar potential:
\begin{eqnarray}\label{2:pot}
\langle V_{\mathrm{neutral}}
\rangle 
&=&
\frac{g_1^2+g_2^2}{8} \left( |\nu|^2 + |v_1|^2 - |v_2|^2 \right)^2
\nonumber \\
&+&
|\lambda|^2 \left(
|\nu^c|^2 |v_1|^2 + |\nu^c|^2 |v_2|^2 + |v_1|^2 |v_2|^2\right) +
|\kappa|^2 |\nu^c|^4 \nonumber \\
& + & 
|Y_{\nu}|^2 \left(
|\nu^c|^2 |v_2|^2 + |\nu^c|^2 |\nu|^2 + |\nu|^2 |v_2|^2 \right)
\nonumber \\
& + &
m_{H_1}^2 |v_1|^2 + m_{H_2}^2 |v_2|^2 + m_{\tilde \nu^c}^2 |\nu^c|^2
+ m_{\tilde \nu}^2 |\nu|^2
\nonumber \\
& + &
\left(-\lambda \kappa^* v_1 v_2 {\nu^c}^{*2}
-\lambda Y_{\nu}^* |\nu^c|^2 v_1  \nu^*
-\lambda Y_{\nu}^* |v_2|^2 v_1  \nu^*
+ k Y_{\nu}^* v_2^* {\nu}^*  {\nu^c}^{2} \right.
\nonumber \\
&-& \left. \lambda A_\lambda \nu^c v_1 v_2 +
Y_{\nu} A_{\nu} \nu^c \nu v_2 
+ \frac{1}{3}
\kappa A_\kappa {\nu^c}^3 + \mathrm{H.c.}
\right) \,.
\end{eqnarray}
In the following, we assume for simplicity that all parameters
in the potential are real.
One can derive the four minimization conditions with respect to the VEVs
$v_1$, $v_2$, $\nu^c$, $\nu$, with the result
\begin{eqnarray}\label{2:minima}
\frac{1}{4}(g_1^2+g_2^2)(\nu^2+v_1^2-v_2^2)v_1 + \lambda^2 v_1\left( {\nu^c}^2
  + v_2^2\right)+ m_{H_1}^2 v_1 
-\lambda \nu^c v_2 \left(\kappa \nu^c  +A_\lambda \right) 
\nonumber \\
-\lambda Y_\nu \nu  \left({\nu^c}^2  + v_2^2 \right) 
= 0 \,,
\nonumber \\
-\frac{1}{4}(g_1^2+g_2^2)(\nu^2+v_1^2-v_2^2)v_2 + \lambda^2 v_2\left( {\nu^c}^2
  + v_1^2\right)  + m_{H_2}^2 v_2 
-\lambda \nu^c v_1 \left(\kappa \nu^c  +A_\lambda \right) 
\nonumber \\
+ Y_\nu^2 v_2 \left({\nu^c}^2  + \nu^2 \right) 
+ Y_\nu \nu  \left(-2 \lambda v_1v_2 + \kappa {\nu^c}^2  + A_\nu \nu^c \right) 
= 0 \,,
\nonumber \\
\lambda^2 \left(v_1^2 + v_2^2\right) \nu^c 
+2 \kappa^2 {\nu^c}^3 +
m_{\tilde\nu^c}^2  \nu^c
-2\lambda\kappa v_1 v_2 \nu^c 
-\lambda A_{\lambda} v_1v_2 + \kappa A_{\kappa} {\nu^c}^2
\nonumber \\
+ Y_\nu^2 \nu^c  \left(v_2^2  + \nu^2 \right) 
+ Y_\nu \nu  
\left(-2 \lambda \nu^c v_1 
+ 2\kappa v_2 {\nu^c}  + A_\nu v_2\right) 
=0 \,,
\nonumber \\
\frac{1}{4}(g_1^2+g_2^2)(\nu^2+v_1^2-v_2^2)\nu + m_{\tilde\nu}^2 \nu
\nonumber \\
+ Y_\nu^2 \nu \left( v_2^2 +  {\nu^c}^2   \right)   
+ Y_\nu 
\left(-\lambda {\nu^c}^2 v_1  -\lambda v_2^2 v_1  
+ \kappa v_2 {\nu^c}^2  + A_\nu \nu^c v_2\right) 
= 0 \,.
\end{eqnarray}
As discussed in the context of $R$-parity breaking models with extra
singlets \cite{Masiero},
the VEV of the left-handed sneutrino,
$\nu$, is in general small.
Here we can use the same argument.
Notice that in the last equation in (\ref{2:minima})
$\nu\to 0$ as $Y_{\nu}\to 0$, and since
the coupling $Y_{\nu}$ determines the Dirac mass for the
neutrinos,  $m_D \equiv Y_{\nu}v_2$, $\nu$ has to be very small.
Using this rough argument we can also get an estimate of the value,
$\nu\lsim m_D$.
This also implies that, neglecting terms proportional to $Y_{\nu}$,
we can approximate the other three equations as the ones
defining the minimization conditions for the NMSSM, with the 
substitution ${\nu^c}\leftrightarrow s$.
Thus one can carry out the analysis of the model
similarly to the NMSSM case, where many solutions in the parameter space
$\lambda ,\kappa ,\mu (\equiv\lambda s) , \tan \beta , A_\lambda , A_\kappa$,
can be found (see e.g. ref.~\cite{nmssm2} and references therein).

Once we know that solutions are available in this model,
we have to discuss in some detail the important issue of mass
matrices.
Concerning this point, the breaking of $R$-parity makes the $\mu$$\nu$SSM
very different from MSSM and NMSSM.
In particular, neutral gauginos and Higgsinos are now mixed
with the neutrinos.
Not only the
fermionic component of $\hat \nu^c$ mixes with the neutral Higgsinos
(similarly to the fermionic component of $\hat S$ in the NMSSM), but
also the fermionic component of 
$\hat \nu$ enters in the game,
giving rise to a
sixth state. 
Of course, now we have to be sure that one eigenvalue of this matrix
is very small, reproducing the experimental results about neutrino masses.
The neutral fermion mass matrix is 
\begin{equation}
  \mathcal{M}_{\mathrm{n}} = \left(
    \begin{array}{cc}
M & m \\
m^T & 0 \\
\end{array} \right),
  \label{neumatrix}
\end{equation}
where 
{\footnotesize \begin{equation}
  M
= \left(
    \begin{array}{ccccc}
      M_1 & 0 & 
-M_Z \sin \theta_W \cos \beta 
&
M_Z \sin \theta_W \sin \beta 
& 0  \\
      0 & M_2 & 
M_Z \cos \theta_W \cos \beta 
&
      -M_Z \cos \theta_W \sin \beta 
& 0 \\
      -M_Z \sin \theta_W \cos \beta &
      M_Z \cos \theta_W \cos \beta &
      0 & -\lambda \nu^c & -\lambda v_2 \\
      M_Z \sin \theta_W \sin \beta &
      -M_Z \cos \theta_W \sin \beta &
      -\lambda \nu^c &0  & -\lambda v_1 + Y_\nu \nu \\
      0 & 0 &  -\lambda v_2 & -\lambda v_1+ Y_\nu \nu & 2 \kappa \nu^c
    \end{array} \right),
  \label{neumatrix2}
\end{equation}
}
is very similar to the neutralino mass matrix of the NMSSM 
(substituting  ${\nu^c}\leftrightarrow s$ and neglecting
the contributions $Y_\nu \nu$),
and 
\begin{equation}
 {m}^T
=
\left(\,\,\,\,\,\,  -\frac{g_1\nu}{\sqrt 2}\,\,\,\,\,\,   \frac{g_2\nu}{\sqrt 2}\,\,\,\,\,\,
  0\,\,\,\,\,\,   
Y_\nu\nu^c\,\,\,\,\,\,  
Y_\nu v_2\,\,\,\,\,\,     \right)\ .
  \label{neumatrix3}
\end{equation}

Matrix (\ref{neumatrix}) is a matrix of the see-saw type that will
give rise to a very light eigenvalue if
the entries of the matrix $M$ are much larger than the entries
of the matrix $m$.
This is generically the case since the entries of $M$ are of order the
EW scale, but for the entries of $m$, $\nu$ is small 
and $Y_\nu v_2$ is the Dirac mass for the neutrinos $m_D$ 
as discussed above
($Y_\nu\nu^c$ has the same order of magnitude of $m_D$).
We have checked numerically that correct neutrino masses can easily be
obtained.
For example, using typical EW-scale values in (\ref{neumatrix2}), 
and a Dirac mass of order $10^{-4}$ GeV
in (\ref{neumatrix3}),
one obtains that the lightest eigenvalue of (\ref{neumatrix}) 
is of order $10^{-2}$ eV.
Including the three generations in the analysis we can obtain different
neutrino mass hierarchies playing with the 
hierarchies in the Dirac masses.
It is worth noticing here that 
because of the matrix (\ref{neumatrix2}),
the presence of the 
last type of terms in (\ref{superpotentialmunussm}) 
is not essential to generate Majorana masses for the neutrinos.
Let us finally mention that the neutrino sector in the context of the superpotential (\ref{superpotentialmunussm}) 
has also been considered for one family of right-handed neutrinos in refs.~\cite{pandita,abada}.
In ref.~\cite{moreau}, bilinear terms have been added.

On the other hand, the charginos mix with the charged leptons.
One can check that
there will always be a light eigenvalue corresponding to the 
electron mass $Y_e v_1$. 
The extension of the analysis to three generations is again
straightforward.
 
Of course, other mass matrices are also modified.
This is the case for example of the Higgs boson mass matrices.
The presence of the VEVs $\nu$, $\nu^c$, leads to mixing of the
neutral Higgses with the sneutrinos. 
Likewise the
charged Higgses will be mixed with the charged sleptons.
On the other hand,
when compared to the MSSM case, the structure of squark mass terms
is essentially unaffected, provided that one uses $\mu = \lambda\nu^c
$,
and neglects the contribution of the fourth term 
in (\ref{superpotentialmunussm}).



\section{Conclusions}

We are all very lucky to live in a historic moment for particle
physics and science in general,
the moment when the LHC is switched on.
This huge machine, under construction for more than eight years,
will finally start operations at the end of this year.
The LHC will be able to answer not only a crucial question such as
the origin of the mass, but also to clarify whether or not
a new symmetry in Nature with 
spectacular experimental implications exists.
Of course, 
if the Higgs is finally found, this will be a great success for 
everybody,
but let us be more ambitious and hope that also one of the SUSY models 
described here will be confirmed.

\section*{Acknowledgments}
We gratefully acknowledge D.E. L\'opez-Fogliani for very useful comments.
This work was supported 
in part by the Spanish DGI of the
MEC under Proyectos Nacionales FPA2006-05423 and FPA2006-01105;
by the Comunidad de Madrid under Proyecto HEPHACOS,
Ayudas de I+D S-0505/ESP-0346; and
also by the European Union under the RTN programs  
MRTN-CT-2004-503369 and HPRN-CT-2006-035863, 
and under the ENTApP Network of the ILIAS
project RII3-CT-2004-506222.


\end{document}